\documentclass[paper]{agujournal2019}
\usepackage{url}
\usepackage{lineno}
\usepackage{color}
\usepackage{booktabs}
\usepackage{amsmath}
\usepackage{amssymb}
\usepackage{graphicx}
\usepackage[export]{adjustbox}
\usepackage{diagbox}
\usepackage{multirow}
\draftfalse

\newcommand{\planss} {{Planatary Space Science }}  
\newcommand{\ssr}{   {Space Sci. Rev. }}

\newcommand{\jgr}{   {J. Geophys. Res.}}
\newcommand{\grl}{   {Geophys. Res. Lett.}}
\newcommand{\apj}{   {Astrophys. J.}}
\newcommand{\apjl}{   {Astrophys. J. Lett.}}

\newcommand{\nat}{   {Nature}}
\newcommand{\aap}{   {Astronomy \& Astrophysics}}

\journalname{JGR: Space Physics}

\begin{document}


\title{Upper Limit of Electron Energization in the Near-Earth Plasma Sheet during Substorm Injections}

\authors{Weiqin Sun\affil{1}, Xiao-Jia Zhang \affil{1}, Anton V. Artemyev\affil{2}, Xi Lu \affil{1}, Xinlin Li\affil{3,4}, Yang Mei\affil{3,4}, Zheng Xiang\affil{3}, Declan O'Brien\affil{3,4}}

\affiliation{1}{Department of Physics, University of Texas at Dallas, Richardson, Texas, USA}
\affiliation{2}{Department of Earth, Planetary, and Space Sciences, University of California, Los Angeles, Los Angeles, California, USA}
\affiliation{3}{Laboratory for Atmospheric and Space Physics, University of Colorado Boulder, Boulder, CO, USA}
\affiliation{4}{Department of Aerospace Engineering Sciences, University of Colorado Boulder, Boulder, CO, USA}


\correspondingauthor{Sun Weiqin}{weiqin.sun@utdallas.edu}

\begin{keypoints}
\item CIRBE observations in substorm magnetotail suggest the formation of relativistic electrons well tailward of the outer radiation belt
\item RCM simulations confirm CIRBE observations and reproduce the spectra of $250-3000$keV electrons observed in the magnetotail
\item RCM simulations suggest that substorm magnetic field reconfiguration can adiabatically accelerate electrons to $1-3$MeV
\end{keypoints}

\begin{abstract}
The Earth's magnetotail, located on the night side of the magnetosphere, is a dynamic region where magnetic field energy is released and converted into plasma heating, particle acceleration, and kinetic energy through magnetic reconnection. Recent low-altitude observations from the CIRBE CubeSat reveal that the efficiency of particle acceleration in the magnetotail can be high enough to produce relativistic and ultra-relativistic electrons with energies reaching several MeV.
To investigate the underlying acceleration mechanisms, we used the Rice Convection Model (RCM) to simulate the observed magnetotail electron populations. The simulations successfully reproduced key features of CIRBE observations, including the spectral shape and energy range of accelerated electrons. This agreement between RCM results and CIRBE observations offers crucial insights into the physical processes responsible for extreme electron acceleration events in the magnetotail.
\end{abstract}

\section{Introduction}
Sudden and significant enhancements of relativistic electrons in the inner magnetosphere, especially in the outer radiation belt, are closely associated with charged particle acceleration mechanisms in the nightside magnetosphere (magnetotail). These mechanisms play a crucial role in transporting energy flux into the inner magnetosphere and the formation of the ring current and outer radiation belt \cite<see examples in>{Angelopoulos20, Lin21, Cohen21, Hua23, Sorathia23, Michael24:wave&gamera}. In Earth's magnetotail, charged particle acceleration is primarily driven by magnetotail reconnection, a dominant and well-documented process in the near-Earth space \cite{Baker96, Angelopoulos08, Paschmann13,Sega&Ergun24}. In contrast to local electron acceleration in the radiation belts via repeated gyro-resonant interactions with whistler-mode chorus waves \cite<see examples in>{Horne05Nature, Summers98, Thorne13:nature}, magnetotail acceleration typically involves electron transport into regions of stronger magnetic field, i.e., these are adiabatic energization processes \cite<e.g.,>{Birn12:SSR,Birn22:electrons,Gabrielse19,Eshetu19}. 
These earthward injections from the magnetotail, typically observed as rapid flux increases at radial distances of 5 to 9 \( R_E \) \cite{Friedel&Korth96, Reeves90,Birn98,Li03:injection,Gabrielse14}, are commonly associated with substorm dipolarization -- a sudden reconfiguration of the nightside magnetosphere from tail-like to more dipolar \cite<e.g.,>{Runov09grl,Runov11jgr,Liu11:DF}. In particular, substorm injections provide a key source of energetic electrons in the inner magnetosphere, which are then further accelerated through diffusive inward radial transport \cite<driven by fluctuating electric and magnetic fields; see examples in>{Brautigam&Albert00, Shprits08:JASTP_transport,Millan&Baker12} and local wave–particle interactions by chorus waves \cite<see examples in>{Shprits08:JASTP_local,Thorne21:AGU}. Together, magnetotail acceleration and subsequent injections play a crucial role in populating the inner magnetosphere and increasing electron fluxes at geosynchronous orbit and closer to Earth.

Over the past two decades, observations and modeling have consistently shown that substorms can promptly inject both subrelativistic (10--100 keV) and relativistic (\(\geq\)MeV) electrons into the inner magnetosphere \cite<e.g.,>{Turner15, Dai15, Turner21, Kim23, califf22}, significantly contributing to radiation belt enhancements. For instance, GPS and geosynchronous spacecraft observations reveal that strong substorm-associated inductive electric fields (\(\gtrsim10\) mV/m) can transport electrons across a broad energy range to geosynchronous orbit, encompassing typical substorm electrons (50–300 keV) and relativistic electrons (300keV to several MeV). This provides a unidirectional transport mechanism supplementing radial diffusion and leading to exceptionally strong relativistic electron fluxes during substorms \cite{Ingraham01}. Akebono observations similarly show rapid enhancements of \(>\)2.5 MeV electrons in the outer belt during storm-time substorm dipolarizations \cite{Nagai06}. Data from Cluster, Polar, LANL, GOES, and Van Allen Probes have confirmed that substorm dipolarization electric fields inject MeV electrons into the outer radiation belt by transporting magnetotail electrons to geosynchronous orbit, driving flux enhancements on substorm timescales \cite{Dai14, Dai15}. Van Allen Probes and MMS further demonstrate that suchinjections can penetrate to low $L$-shells (as low as \( L \approx 2.5 \)) with energies up to $250$ keV\cite{Turner15}, and that the magnetotail plasma sheet can be a rapid, localized source of $>1$ MeV electrons (to the outer radiation belt) during active times \cite{Turner21}. During CIR-driven geomagnetic storms, successive dipolarizations lead to rapid (within hours) injections of 100 keV–MeV electrons, resulting in a prompt  increase in relativistic electron phase space density (PSD) by factors of 4–10, primarily through near-equatorial betatron acceleration \cite{Xiong22}.

Recent statistical analysis of THEMIS observations (2015–2019) reveal that turbulent electric fields generated during bursty bulk flow (BBF) braking in the magnetotail can locally energize electrons, producing a pre-accelerated population that may feed into the outer radiation belt \cite{Usanova22}. MMS has provided high-resolution insights into particle energization processes within the magnetotail: \citeA{Ergun20:observations} showed intense plasma heating and particle acceleration via localized (turbulent) electric fields  generated by magnetic reconnection. More recently, a rare MMS-observed electron diffusion region (EDR) in the turbulent magnetotail has shown strong energy transport and particle acceleration through a runaway-like process, driven by low-density inflow and positive feedback between turbulence and reconnection \cite{Qi24}. MMS measurements also show that electrons below 200 keV can be significantly energized through Fermi and betatron processes around traveling flux ropes and reconnection X-lines, reaching MeV energies in flux rope cores through multidimensional and turbulent magnetic field effects \cite{Sun22}. In addition, MMS has captured direct acceleration of relativistic electrons (80–560 keV) at the reconnection X-line, with enhanced fluxes and distinct spectral features observed in the separatrix regions \cite{Sun25}. Together, these observations highlight a consistent picture of localized and efficient electron acceleration in the magnetotail, driven by the coupled action of BBF-associated turbulence and reconnection-driven processes across multiple spatial and temporal scales.

These observational findings are further supported by a range of modeling studies. Test particle tracing in MHD model fields \cite{Birn&Hesse96} during the substorm dipolarization have shown that tens of keV plasma sheet electrons can be transported earthward from \( x \approx -20 R_E \) to \( x \approx -10 R_E \), gaining an order of magnitude in energy through \( E \times B \) drift in the dipolarization region, and potentially reaching MeV energies if further transported inward while conserving the first adiabatic invariant \cite{Kim00}.  A three-dimensional electron kinetic model, incorporating convection and radial diffusion, reveals the essential role of inductive electric fields (associated with magnetic reconfiguration) in facilitating such transport and producing large electron flux enhancements around geosynchronous orbit \cite{Fok01}.  Additional modeling efforts---using a kinetic model of the radiation belt coupled with MHD fields \cite{Glocer09}---demonstrate rapid enhancement of relativistic electron fluxes due to dipolarization, in agreement with Akebono observations \cite{Glocer11}.
Likewise, a newly developed radiation belt model, which integrates test particle tracing and coupled 3D ring current–MHD simulations driven by solar wind data, effectively captures the dramatic fluctuations in outer radiation belt electrons, as verified by Van Allen Probe measurements \cite{Sorathia18}.

More recently, both targeted and data-driven simulations have further illuminated the nature of localized dipolarization and injections. For example, analytical modeling of dipolarizing flux bundles (DFBs) shows that sharp magnetic and electric field gradients can efficiently transport and energize electrons, producing spatially localized injection signatures that depend sensitively on spacecraft location and front structure \cite{Gabrielse16}. MHD simulations of bursty bulk flows (BBFs) also reveal that electrons can be injected and accelerated from beyond \( -24 R_E \) into geosynchronous orbit, with energization strongly dependent on initial pitch angle, location, and nonadiabatic effects \cite{Eshetu19}. A global test-particle model coupled with the MAGE geospace simulation demonstrates that resonant interactions with lower-band chorus waves can rapidly accelerate electrons to relativistic energies during geomagnetic storms, with strong spatial variations shaped by magnetic and plasma structures \cite{Michael24:wave&gamera}.  Finally, both test-particle simulations and MMS observations show that mesoscale dipolarization flows can generate strong parallel and perpendicular electron temperature anisotropies, leading to kinetic instabilities and intense wave activity (e.g., whistler-mode waves, broadband emissions) in the magnetotail plasma  \cite{Ukhorskiy22:NatSR}.

However, the transient nature of electron acceleration in the magnetotail, combined with the energy limitations of energetic particle detectors on magnetotail missions—typically lacking high-resolution measurements at relativistic energies \cite{Wilken01, Angelopoulos08:sst, Blake16}—makes it challenging to track electron injection and acceleration process from mid-tail reconnection sites to the inner magnetosphere using near-equatorial observations alone \cite<see discussion in>{Shumko24:arXiv,Zhang25:ELFIN&RX}. A valuable alternative is provided by low-altitude, polar-orbiting spacecraft \cite<e.g.,>{Wing&Newell02,Sergeev19,Sergeev23:elfin}, which traverse the full radial extent of the near-Earth magnetotail in minutes and offer high-resolution snapshots of equatorial electron flux dynamics \cite<see discussion in>{Artemyev22:jgr:ELFIN&THEMIS}.

In this study, we analyze several substorm events where relativistic (up to several MeV) electrons were observed in the near-Earth plasma sheet by the Colorado Inner Radiation Belt Experiment (CIRBE) CubeSat \cite{Li2022,Li24:grl:CIRBE}. To investigate the characteristics of the electron populations accelerated in the midtail, we perform numerical simulations using the Rice Convection Model (RCM) and compare the simulated energetic electron fluxes with CIRBE measurements during substorms.

The RCM is a well-established, first-principles model of Earth's magnetosphere \cite<see>[and references therein]{Toffoletto03}. It calculates both \( E \times B \) and gradient/curvature drift velocities for isotropic, \(\lambda\)-conserving particles \cite{Wolf83}. Here, \(\lambda_k\) is the energy invariant, conserved as particles drift within flux tubes filled with isotropic particles of kinetic energy \(W_k\) \cite{Wolf83,Schulz&Chen08}. The electric \( E \) and magnetic \( B \) fields are self-consistently calculated \cite{Toffoletto03,Yang19}, with the electric potential derived from the current continuity between the magnetosphere and the ionosphere. The magnetic field profiles are self-consistently calculated using a finite-volume MHD code integrated with RCM \cite{Silin13,Yang19}. This makes RCM particularly suitable for simulating the transport of energetic electrons in the midtail, where these particles are usually isotropic \cite{Artemyev22:jgr:ELFIN&THEMIS} and follow both \( E \times B \) and gradient/curvature drifts \cite<e.g.,>{Birn14,Gabrielse19}. Although RCM's isotropic assumption limits resolution of anisotropic dynamics, RCM remains valuable for exploring substorm-driven electron transport, complementing low-altitude observations. While relativistic electrons are not present at the start of the simulation, they naturally emerge in the RCM through adiabatic heating—where inward transport during dipolarization compresses flux tubes and shortens drift paths, leading to significant increases in particle energy. This process energizes electrons injected from the RCM outer boundary to relativistic levels as the substorm evolves.

This study aims to combine RCM simulations with CIRBE data to identify key characteristics of electron acceleration following substorm onset. By modeling the substorm expansion phase and comparing the simulated evolution of low-altitude electron fluxes with CIRBE observations, we seek to unravel the potential mechanism of low-altitude energetic electron acceleration. The paper is structured as follows: Section~\ref{sec:data} describes the CIRBE datasets and instrument used for measuring energetic electrons, along with representative substorm injection events observed by CIRBE. Section~\ref{sec:model} presents the RCM simulations constrained by CIRBE observations from Section~\ref{sec:data}, and includes a direct comparison between simulation results and CIRBE measurements. Finally, Section~\ref{sec:discussion} interprets the findings in the context of substorm dynamics and summarizes implications for the origin of these processes.

\section{Dataset}\label{sec:data}
CIRBE is a 3U CubeSat equipped with the Relativistic Electron and Proton Telescope Integrated Little Experiment-2 (REPTile-2) \cite{Khoo22,Li2022}, designed to measure electrons in the 250–6000 keV energy range across sixty energy channels with a 1-second time resolution \cite{Li24:grl:CIRBE}. With a $51^\circ$ field-of-view and a look direction nearly perpendicular to the background magnetic field, REPTile-2 is optimized for measuring perpendicular ($90^\circ$) fluxes, which correspond to the locally trapped flux \( j_{\text{trap}} \) \cite{Zhangkun20}. CIRBE provides two key advantages for studying dynamics of energetic electron populations in the magnetotail: (1) high telemetry efficiency, ensuring near-continuous data coverage from nearly every orbit ($\sim90$ min), and (2) superior energy resolution and sensitivity, enabling detailed spectral measurements of these populations.

In this study, CIRBE measurements are compared with the equatorial THEMIS measurements \cite{Angelopoulos08:ssr} to aid in identifying details of magnetotail current sheet reconfiguration associated with different substorm phases\cite{Runov21:angeo}. We analyze CIRBE observations during multiple substorms, when energetic (and even relativistic) electron injections were detected. While CIRBE measurements at a given location can suggest whether the electrons are most likely trapped, quasi-trapped, or precipitating, this inference relies heavily on the measurement location. A more robust identification of the plasma sheet–outer radiation belt transition, i.e., the so-called isotropy boundary (IB) \cite{Sergeev83,Dubyagin02}, can be achieved by complementing CIRBE data with precipitating and trapped electron flux measurements from nearby POES/MetOp satellites \cite{Evans&Greer04}.

\begin{figure}[!htbp]
    \centering
    \includegraphics[width=1\linewidth]{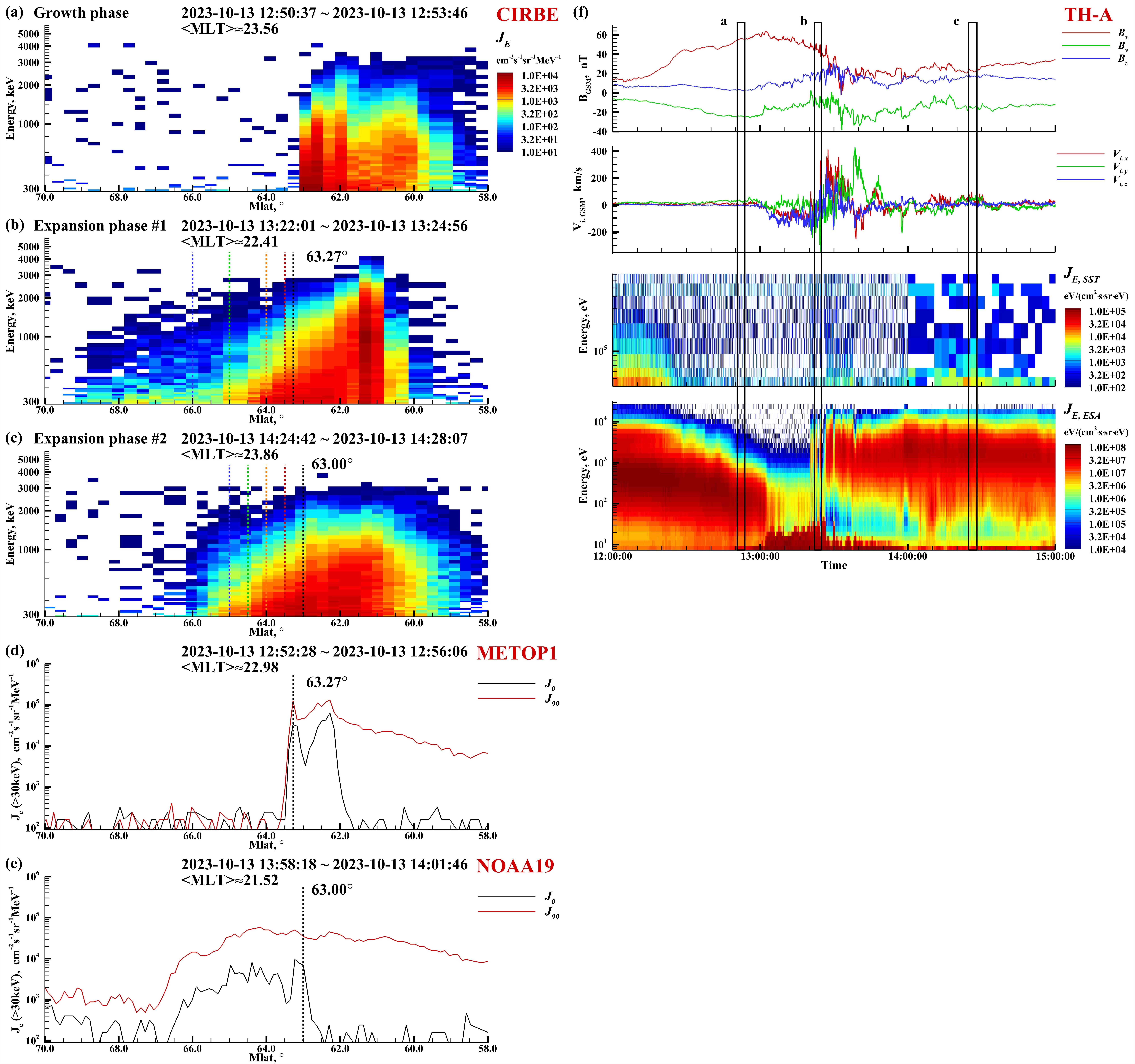}
    \caption{Overview of the first relativistic electron injection event on October 13, 2023. Panels (a)–(c) show the continuous CIRBE observations covering the outer radiation belt and its tailward extension in the nightside magnetosphere (the same CIRBE measurements presented in the energy-time domain can be found in the Supplementary Information). In Panels (b) and (c), the black dashed lines represent the electron isotropy boundaries determined from nearby POES satellite observations, while the colored dashed lines indicate the tailward positions of these boundaries, where flux-energy profiles were extracted for subsequent comparison with simulation results. Panels (d) and (e) show the precipitating and trapped electron fluxes (\(>\)30 keV) measured by METOP1 and NOAA19, respectively, with the electron isotropy boundaries indicated by black dashed lines. Panel (f) shows the corresponding THEMIS-A observations, where CIRBE observations are marked by black solid rectangles. From top to bottom, it shows magnetic field measurements from the THEMIS Fluxgate Magnetometer, plasma flow speed and energetic electron fluxes recorded by the Electrostatic Analyzer (ESA) and the Solid State Telescope (SST).}
    \label{fig1}
\end{figure}

Figure \ref{fig1} presents a typical relativistic electron injection event during a substorm on 13 October, 2023. Figs. \ref{fig1}a-\ref{fig1}c show three CIRBE orbits with continuous observations covering the outer radiation belt and its tailward extension in the nightside magnetosphere. The corresponding time and the average magnetic local time (MLT) are indicated at the top.
Fig. \ref{fig1}f shows THEMIS-A observations for the same time period, with CIRBE observation intervals marked by solid black rectangles and labeled with the corresponding subplot letters. The three-hour THEMIS-A observations reveal distinct magnetic field and plasma variations characteristic of substorm events. Between \(T = 12:00:00\) UT and \(T = 13:00:00\) UT, magnetic field measurements from the THEMIS Fluxgate Magnetometer \cite{Auster08:THEMIS} show a gradual decrease in \( B_z \) and an increase in \( B_x \); this is a typical substorm growth phase signature, as a result of current sheet thinning and magnetic field-line stretching \cite<see>{Petrukovich07,Artemyev16:jgr:thinning,Yushkov21}. After \(T = 13:00:00\) UT, \( B_z \) begins to rise while \( B_x \) decreases, marking the transition toward dipolarization after substorm onset \cite<see>{Sitnov19,Runov21:jastp}. Around \(T = 13:20:00\) UT, plasma flow speed \cite{McFadden08:THEMIS} exhibits a strong earthward enhancement, accompanied by a significant increase in energetic electron fluxes \cite{Angelopoulos08:ssr}, as observed by the Electrostatic Analyzer (ESA) and the Solid State Telescope (SST); these are typical signatures of plasma injections and fast plasma flows arriving from tailward reconnection region\cite<e.g.,>{Runov09grl,Hwang11,Gabrielse14}. These THEMIS observations indicate that the substorm onset likely occurred shortly after \(T = 13:00:00\) UT. Based on this timing, we associate CIRBE observations in Fig. \ref{fig1}a–\ref{fig1}c with the growth phase, early expansion phase (immediately after substorm onset), and late expansion phase, respectively (note, this classification relies solely on THEMIS data; ground-based magnetic or optical observations are not used here). The electron spectra in Fig. \ref{fig1}b and \ref{fig1}c are further analyzed in conjunction with observations from METOP1 (Fig. \ref{fig1}d) and METOP3 (Fig. \ref{fig1}e). To select the most relevant POES/MetOp satellites (from NOAA-15, NOAA-18, NOAA-19, METOP1, METOP2, and METOP3), we require them to be within $\pm1$ hour in UT and $\pm2$ in MLT, prioritizing the closest in time and location from CIRBE observations. Based on the precipitating and trapped electron fluxes (\(>\)30 keV) measured by METOP1 (Fig. \ref{fig1}d), the electron isotropy boundary (IB) is estimated to be $\sim63.27^\circ$ around $\sim 13$UT (after substorm onset), as indicated by the dashed black lines in Figs. \ref{fig1}b-\ref{fig1}e (for both CIRBE and POES, we use AACGM magnetic latitudes). In $\sim 14$UT, the strong dipolarization suppresses the formation of IB \cite<see discussion in>{Sergeev19}, and NOAA19 observed significantly lower precipitating fluxes in comparison with trapped fluxes (see Fig. \ref{fig1}e). The colored dashed lines in Figs. \ref{fig1}b and \ref{fig1}c indicate the tailward positions of these boundaries, where flux-energy profiles were extracted for subsequent comparison with simulation results (under the assumption that CIRBE measurements tailward (poleward) of the isotropy boundary correspond to the magnetotail with isotropic electron fluxes). Therefore, as shown in Figs. \ref{fig1}b and Fig. \ref{fig1}c, the injected relativistic electrons were observed poleward of the electron isotropy boundary, with energies exceeding 2 MeV. As noted earlier in Fig. \ref{fig1}f, this relativistic electron burst was detected while THEMIS-A also observed enhanced energetic electron fluxes after approximately \(T = 13:20:00\) UT in the nightside magnetosphere. But this flux increase was mainly restricted to below 100 keV. This THEMIS-A and CIRBE comparison underlines the importance of low-altitude monitoring: CIRBE captured localized relativistic electron injections that were not necessarily resolvable by the near-equatorial THEMIS spacecraft \cite<see discussion in>{Zhang25:ELFIN&RX}.

\begin{figure}[!htbp]
    \centering
    \includegraphics[width=1\linewidth]{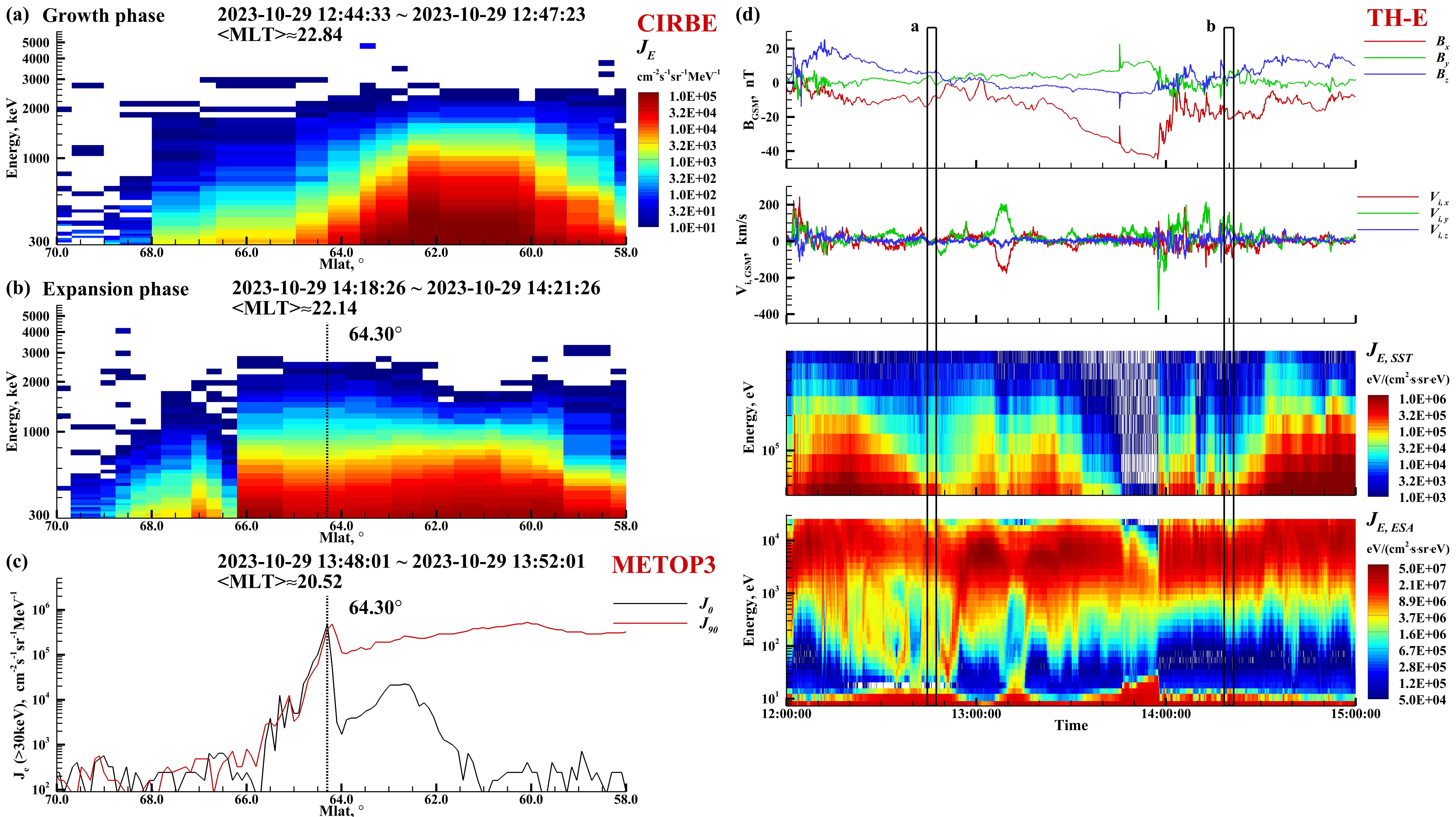}
    \caption{Overview of the second relativistic electron injection event, organized in the same format as Figure \ref{fig1}. (The same CIRBE measurements from panels (a, b) presented in the energy-time domain can be found in the Supplementary Information.)}
    \label{fig2}
\end{figure}

Figure \ref{fig2} shows another relativistic electron injection event. Figs. \ref{fig2}a and \ref{fig2}b show CIRBE observations spanning the outer radiation belt and its tailward extension into the nightside magnetosphere. Concurrently, THEMIS-E observations (Fig. \ref{fig2}d) capture magnetic field and plasma variations indicative of substorm activity.  Following approximately \(T = 14:00:00\) UT, a noticeable increase in \( B_z \) and a strong earthward plasma flow enhancement signal the onset of dipolarization. Simultaneously, ESA and SST data reveal a significant rise in energetic electron fluxes, further confirming the substorm activity \cite<see>{Gabrielse14,Runov15}. Based on THEMIS observations, we associate the CIRBE observations in Fig. \ref{fig2}a with the growth phase, and those in Fig. \ref{fig2}b with the expansion phase. To further investigate the injected electron population, we analyze the electron spectrum in Fig. \ref{fig2}b alongside POES satellite measurements. Precipitating and trapped electron fluxes (\(>\)30 keV) from METOP3 (Fig. \ref{fig2}c) suggest that the electron isotropy boundary (IB) is located around $\sim64.30^\circ$, as indicated by the black dashed lines in Figs. \ref{fig2}b and \ref{fig2}c. Notably, as shown in Fig. \ref{fig2}b, the injected relativistic electrons appear poleward of this boundary, reaching energies exceeding 1 MeV and the spatial structure of these injected electrons ($>66.2^\circ$) is clearly distinct from the outer radiation belt region. Although the IB location inferred from CIRBE data suggests it might lie at $>66.2^\circ$, the relatively coarse magnetic latitude resolution of CIRBE observations in Fig. \ref{fig2}b (where one grid cell may span over $1^\circ$) suggests that the discrepancy between the visually inferred IB from Fig. \ref{fig2}b and the IB determined from POES data remains within a reasonable margin. Unlike the previous event, THEMIS-E observations after approximately \(T = 14:00:00\) UT (Fig. \ref{fig2}d) show not only enhanced energetic electron fluxes but also a significant increase in relativistic fluxes, with energies reaching several hundred keV, highlighting a more pronounced high-energy response in this event \cite<see similar observations in>{Qi24,Sun25,Runov25}.

\begin{figure}[!htbp]
    \centering
    \includegraphics[width=1\linewidth]{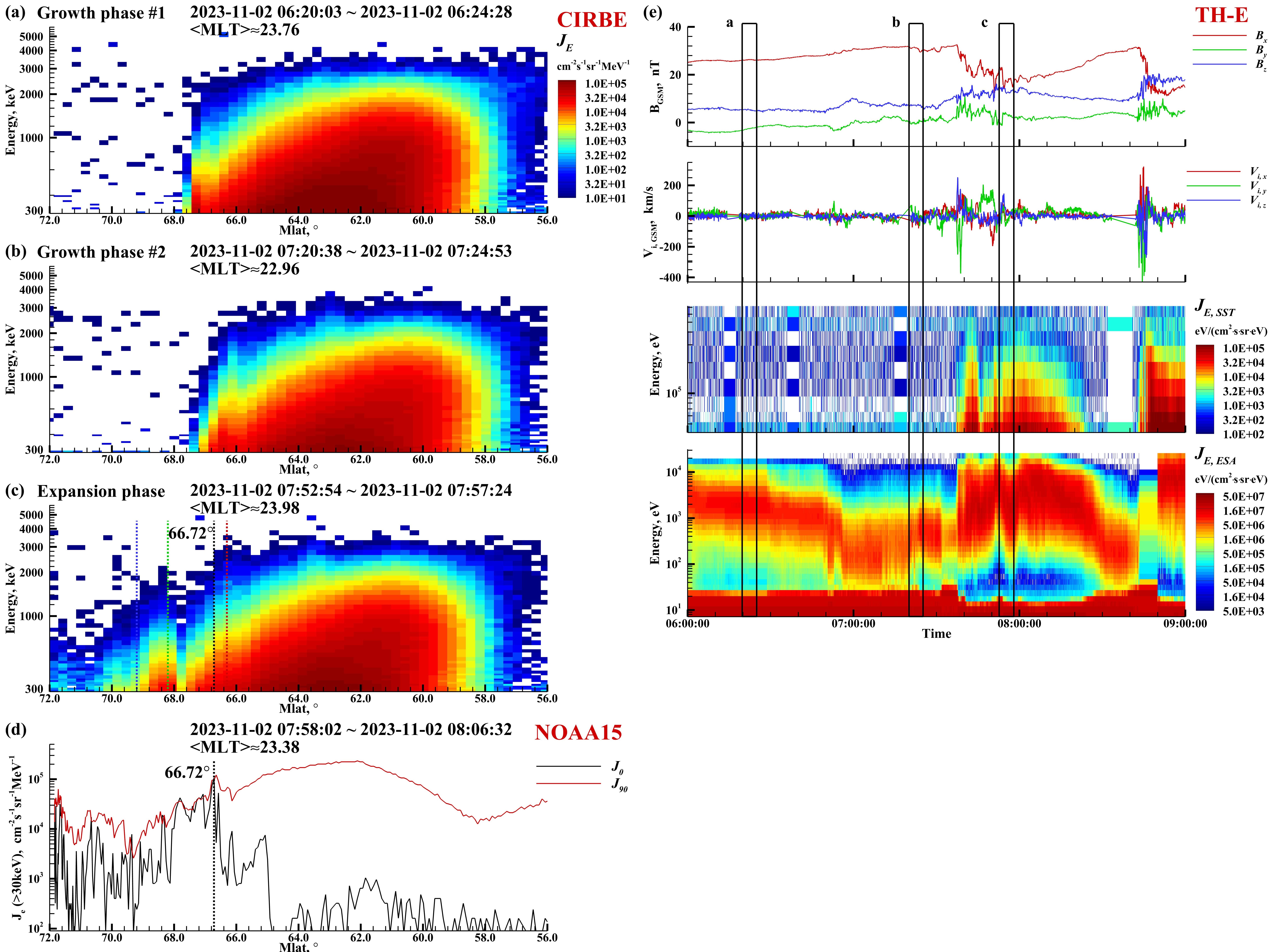}
    \caption{Overview of the third relativistic electron injection event, organized in the same format as Figure \ref{fig1}.(The same CIRBE measurements from panels (a, b) presented in the energy-time domain can be found in SI.)}
    \label{fig3}
\end{figure}

Figure \ref{fig3} presents another relativistic electron injection event, which exhibits a more distinct spatial structure compared to the event in Fig. \ref{fig2}, with the injected electron population clearly separated from the outer radiation belt. CIRBE observations in Figs. \ref{fig3}a–\ref{fig3}c correspond to growth phase 1, growth phase 2 (just prior to the substorm onset), and the expansion phase, respectively, as inferred from THEMIS-E observations in Fig. \ref{fig3}e. Precipitating and trapped electron fluxes (\(>\)30 keV) from NOAA-15 (Fig. \ref{fig3}d) indicate that the electron isotropy boundary (IB) is located around $\sim66.72^\circ$, as denoted by the black dashed lines in Figs. \ref{fig3}c and \ref{fig3}d. Throughout these CIRBE observations, the outer radiation belt structure remains remarkably stable, with its corresponding energy range, magnetic latitude, and flux profile showing minimal variation before and after the substorm. However, in Fig. \ref{fig3}c, a newly injected population of relativistic electrons exceeding 1 MeV emerges, directly linked to substorm-driven injection. Around \(T = 07:40:00\) UT, THEMIS-E observations show a distinct decrease in \( B_x \), an increase in \( B_z \), and an enhanced earthward plasma flow, marking the onset of dipolarization. Simultaneously, ESA and SST measurements record a substantial increase in relativistic electron fluxes, with energies reaching several hundred keV \cite<see similar observations in>{Sun25,Runov25}.

These CIRBE events suggest that post-onset relativistic electron enhancements in the near-Earth magnetotail likely originate from reconnection-driven injections. This interpretation is supported by their location relative to the isotropy boundary observed by POES and their spectral agreement with THEMIS measurements. However, the precise  localization of these enhancements remains uncertain due to the lack of direct in-situ (equatorial) measurements. It is still plausible that these enhancements correspond to the poleward edge of the outer radiation belt rather than distinct magnetotail injections. To resolve this ambiguity, numerical simulations are needed to further investigate whether such relativistic electron populations can originate from magnetotail acceleration processes.

\section{RCM Simulation Results and Comparison with CIRBE Observations}\label{sec:model}

In this study, we utilize the Rice Convection Model (RCM), which assumes fully isotropic electron distributions, to simulate key processes occurring in Earth's plasma sheet during the substorm expansion phase. The simulation incorporates self-consistent electric and magnetic fields, with a high-resolution magnetosphere-ionosphere (M-I) coupling scheme that calculates magnetic-field-aligned currents connecting the magnetosphere to the ionosphere. The initial RCM-E simulations of the substorm growth and expansion phases \cite{toffo2000} introduced a force-balanced magnetic field integrated with plasma pressure derived from the RCM. Building on this foundation, further substorm simulations \cite<e.g.,>{Yang09phd} demonstrated that by carefully adjusting tailward plasma boundary conditions, the RCM can accurately reproduce energetic proton flux variations observed at geosynchronous orbit \cite{Yang08}. Further advancements \cite{Zhang09:rice:convection1, Zhang09:rice:convection} refined RCM-based analyses by incorporating modifications to the Tsyganenko model \cite{Tsyganenko89}, enabling the simulation of magnetic field line stretching and dipolarization. The objective of the RCM simulations in this study is to reproduce the energetic electron flux enhancements at low altitudes after substorm onset, ensuring consistency with CIRBE observations (as described in Section~\ref{sec:data}), rather than relying on the exact physical driver of magnetotail injection in the RCM. Instead of focusing on detailed substorm dynamics, our primary requirement for the RCM is to capture key characteristics of injection-related electron fluxes, particularly their magnitude and timing in response to the substorm onset. By accurately reproducing these flux variations, the RCM provides critical insights into the distribution and evolution of injected electrons, complementing observational constraints from CIRBE.

To reproduce the electron flux enhancements observed by CIRBE at low altitudes during the substorm expansion phase, we designed a three-stage typical substorm event simulation. In the first stage, we first conducted a one-hour generic growth phase simulation before initiating the RCM expansion phase simulation. Specifically, we aim to incorporate realistic energy-magnetic latitude spectra of trapped electrons that cover energies well below the CIRBE energy range (because further electron acceleration during injections involves initial tens of keV to form hundreds of keV populations). Therefore, we utilize ELFIN observations \cite<a low-altitude CubeSat mission providing measurements of $[50,6000]$keV electrons; see>{Angelopoulos20:elfin,Angelopoulos23:ssr} to initialize the electron distribution within the corresponding energy and latitude ranges of the RCM prior to current sheet thinning \cite<data are taken from several events in>{Artemyev22:jgr:ELFIN&THEMIS}. For regions or energy ranges not covered by ELFIN measurements (below 50keV), we used empirical models to initialize the thermal electron distributions \cite{Lemon03, Tsyganenko&Mukai03}. Consistent with these approaches \cite{Yang11, Yang14}, our simulations are initialized using the empirical T89 magnetic field model \cite{Tsyganenko89} and plasma distribution models \cite{Lemon03, Tsyganenko&Mukai03}, with solar wind parameters and geomagnetic indices typical of geomagnetically quiet conditions. The cross-polar cap potential drop is set to 40 kV, following equation (12) of \cite{Zhang09:rice:convection}. The first stage is identical for the three CIRBE events in Section ~\ref{sec:data}. In the second stage of the simulation, we introduce a bubble injection at the tailward boundary of the simulation domain. Starting at substorm onset (\(T = 60\) min), a bubble is initiated at approximately \(r \sim 24 R_E\) by gradually and uniformly reducing the flux-tube entropy \(PV^{5/3}\) over 1 minute, where \(P\) represents plasma pressure. The bubble, which is assumed to be centered at midnight, has a width of 2.0 h in local time at the boundary. Since no direct measurements are available along the outer boundary of the RCM to directly constrain the plasma conditions, we iteratively adjusted the magnitude and duration of entropy reduction until achieving quantitative consistency in electron energy and flux magnitude within the plasma sheet region of the three events analyzed in Section~\ref{sec:data}. The low flux-tube entropy is maintained for \(T = 61–81\) min, resembling the sustained injection of low-entropy plasma during the expansion phase. At \(T = 81\) min, \(PV^{5/3}\) gradually recovers to its pre-injection value within 1 minute. In the third stage, following this recovery, all simulation parameters are held constant at their first-stage values until the end of the run at \(T = 120\) min. Throughout all three stages, the magnetic field is re-equilibrated every 1 minute to ensure consistency in the system’s evolution.

\begin{figure}[!htbp]
    \centering
    \includegraphics[width=1\linewidth]{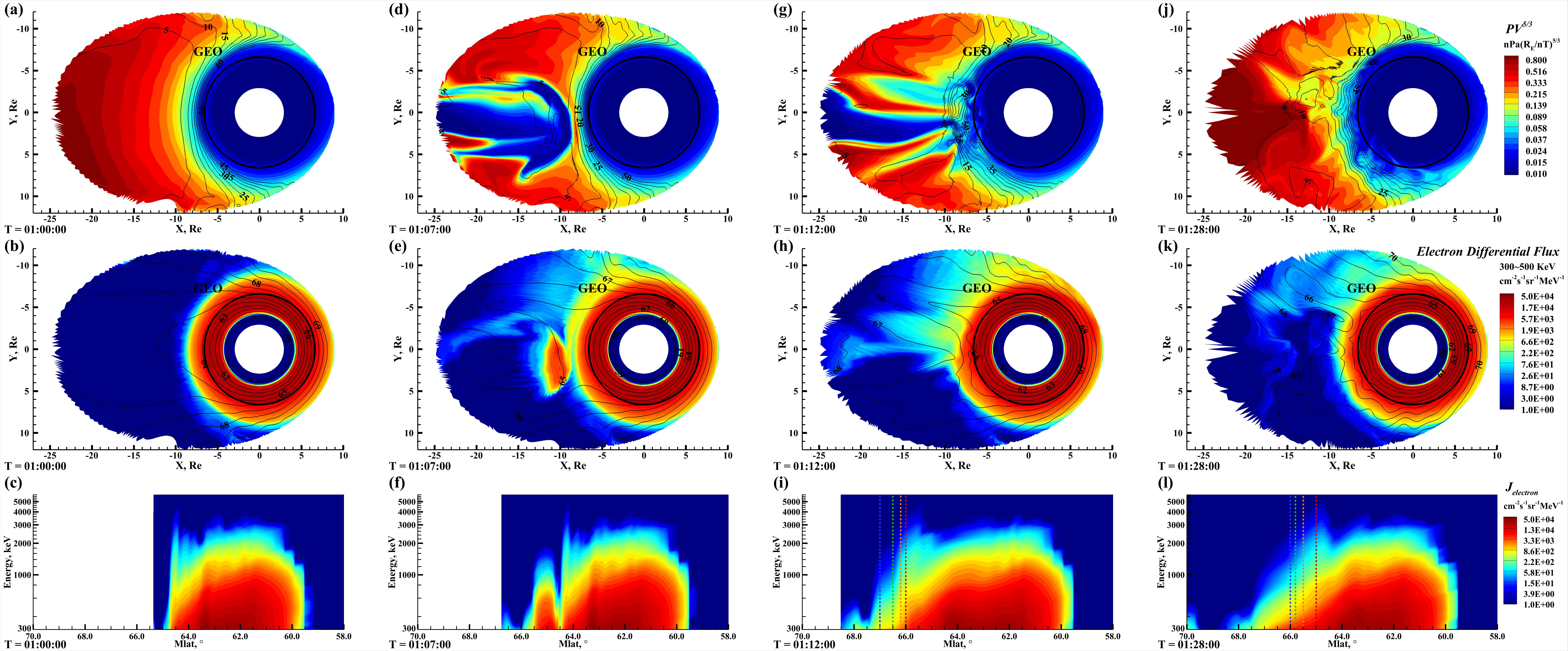}
    \caption{Overview of the RCM expansion-phase simulation results for the CIRBE event in Fig. \ref{fig1}, showing the distribution of flux tube entropy \(PV^{5/3}\), the 300–500 keV electron differential flux in the RCM equatorial plane, and the energetic electron spectrum along the radial direction at midnight at different time snapshots. The black solid lines in Panels (a, d, g, j) and (b, e, h, k) represent equipotential contours of magnetic field strength and magnetic latitude, respectively. The colored dashed lines in Panels (i) and (l) indicate the positions where flux-energy profiles are extracted for subsequent comparison with CIRBE observations.}
    \label{fig4}
\end{figure}

Figure \ref{fig4} presents an overview of the RCM expansion-phase simulation results for the first CIRBE event (from Fig. \ref{fig1}) analyzed in Section~\ref{sec:data}, following a 1-hour substorm growth phase simulation. Figs. \ref{fig4}a-\ref{fig4}c show the distribution of flux tube entropy \(PV^{5/3}\), the 300–500 keV electron differential flux in the RCM equatorial plane, and the energetic electron spectrum along the radial direction at midnight at the end of the growth phase simulation. The black solid lines in Figs. \ref{fig4}a and \ref{fig4}c represent equipotential contours of magnetic field strength and magnetic latitude, respectively. Their corresponding values are labeled in the figures, with units of \(nT\) for magnetic field strength and degrees for magnetic latitude. In Fig. \ref{fig4}c, the same magnetic latitude and energy range as in Fig. \ref{fig1}a are used; however, the flux scale and colorbar differ. This difference arises because the electron spectrum at the end of the growth phase simulation primarily reflects its initial distribution, derived from the energy-magnetic latitude spectra of trapped electrons obtained from ELFIN observations. Despite no specific modifications in the 1-hour substorm growth-phase simulation, the resulting electron spectrum (Fig. \ref{fig4}c) closely resembles the CIRBE spectrum in Fig. \ref{fig1}a. The colorbars in Figs. \ref{fig4}b and \ref{fig4}c are consistent, facilitating a direct comparison between the radial distance in Fig. \ref{fig4}b and the corresponding magnetic latitude in Fig. \ref{fig4}c. Notably, the radial range from \(X \approx 8.5 R_E\) to the RCM outer boundary at \(X \approx 24 R_E\) in Fig. \ref{fig4}b corresponds to only $\sim0.5^\circ$ in magnetic latitude in Fig. \ref{fig4}c. This contraction reflects plasma sheet thinning during the substorm growth phase, resulting in a reduced magnetic latitude range for the projection of the plasma sheet (energetic electron component) to low altitudes\cite<see>{Artemyev22:jgr:ELFIN&THEMIS}. The subsequent simulations for other events in Section~\ref{sec:data} are also initialized based on the results shown in Figs. \ref{fig4}a–\ref{fig4}c. As previously mentioned, the magnitude and duration of entropy reduction were iteratively adjusted to achieve quantitative consistency with each of the three events in Section~\ref{sec:data}. For the event in Fig. \ref{fig1}, the flux-tube entropy \(PV^{5/3}\) is gradually and uniformly reduced to \( 1\% \) of its pre-onset value within 1 minute. Figure \ref{fig4}d shows the low-entropy injection channel in the RCM equatorial plane, which develops alongside magnetic field dipolarization. By \(T = 67\) min, the leading edge of the injected plasma has propagated from the RCM outer boundary to approximately \(X \sim 8 R_E\). At the same location, Fig. \ref{fig4}e shows a significant increase in the 300–500 keV electron differential flux between \(X \sim 8.5 R_E\) and \(X \sim 12 R_E\), a direct consequence of the injection. Notably, the deepest-penetrating electrons have already begun to drift eastward toward the dawnside. In Fig. \ref{fig4}f, the corresponding spectrum clearly exhibits an energetic electron flux burst, with substantial flux enhancements extending up to \(\sim 2\) MeV, aligning in latitude with the flux enhancement region shown in Fig. \ref{fig4}e. As shown in Movie S1, from \(T = 61\) min to \(T = 67\) min, the flux enhancement caused by the bubble injection originates from the RCM outer boundary at high magnetic latitude, gradually propagates toward the burst region in Fig. \ref{fig4}f, and undergoes adiabatic heating, leading to a continuous increase in electron energies, eventually reaching \(\sim 2\) MeV. Meanwhile, the dipolarization of the magnetic field induces a process opposite to plasma sheet thinning during the substorm growth phase, resulting in an expansion of the magnetic latitude range of the plasma sheet (energetic electron component). Figure \ref{fig4}g shows the snapshot of the low-entropy bubble injection inside geosynchronous orbit (GEO), reaching approximately \(X \sim 5 R_E\) by \(T = 72\) min. This inward transport of the bubble continues to influence the electron population in the near-Earth magnetotail. In Fig. \ref{fig4}i, the substantial flux enhancement resulting from the injection merges with the pre-existing high-flux region of the near-Earth magnetotail and down to the inner magnetosphere, forming a new, continuous electron spectrum that extends from the outer radiation belt to the RCM outer boundary. Meanwhile, as dipolarization progresses, the magnetic latitude range of the electron spectrum expands. Figures \ref{fig4}j–\ref{fig4}l capture the evolution of the system a few minutes after the sustained injection of low-entropy plasma ceases. In Fig. \ref{fig4}l, the electron spectrum now spans a broader magnetic latitude range, reflecting the redistribution of energetic electrons following the injection process and reversing the plasma sheet thinning effect that occurred during the substorm growth phase.

\begin{figure}[!htbp]
    \centering
    \includegraphics[width=1.0\linewidth]{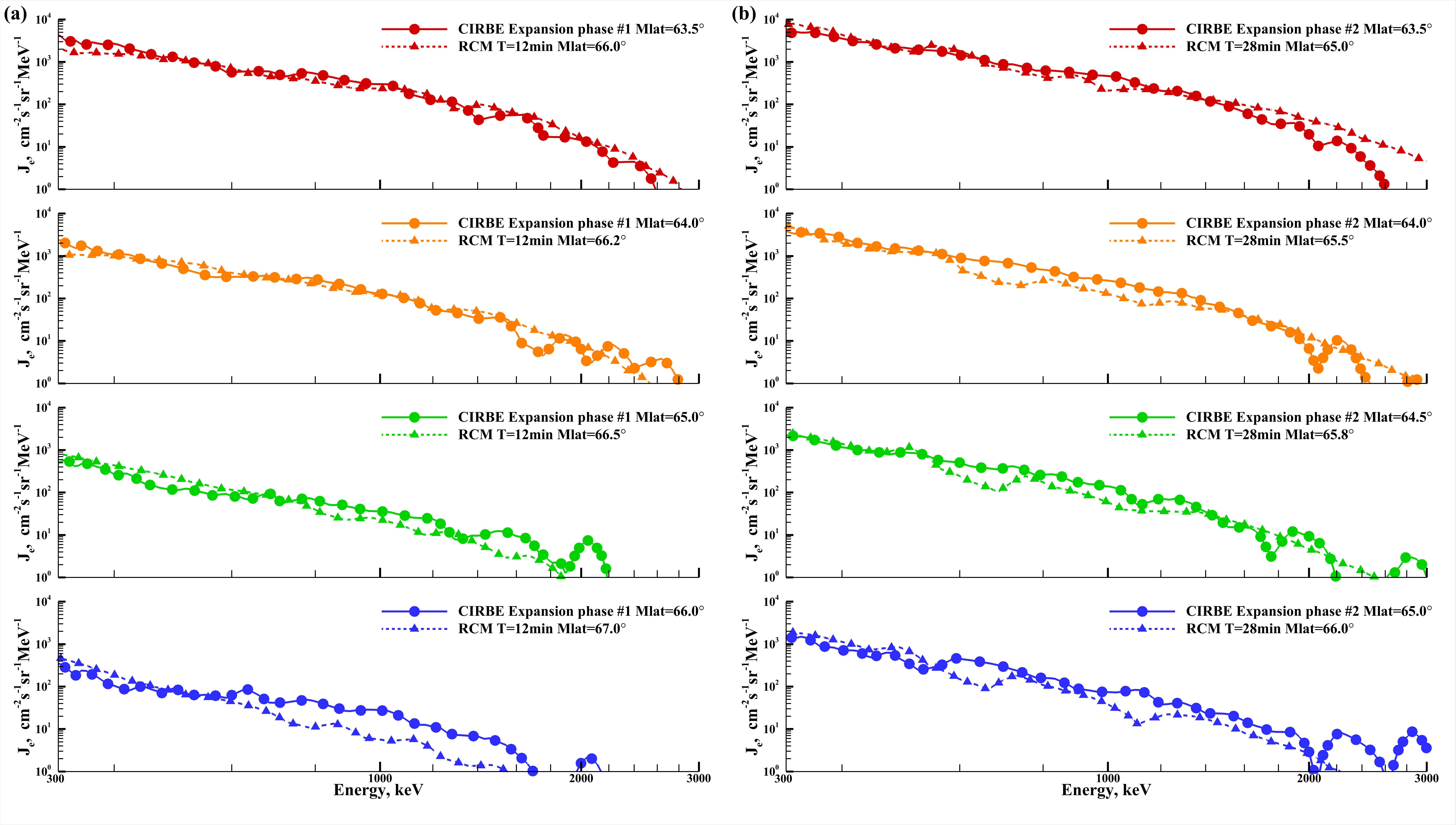}
    \caption{Comparison of the RCM-simulated electron flux with CIRBE observations from Fig. \ref{fig1} during the substorm expansion phase. Panel (a) shows the comparison during the early expansion phase, using the RCM snapshot (Fig. \ref{fig4}i) corresponding to Fig. \ref{fig1}b. Panel (b) shows the comparison during the late expansion phase, using the RCM snapshot (Fig. \ref{fig4}l) corresponding to Fig. \ref{fig1}c. Solid lines with circles represent CIRBE observations, while dashed lines with triangles denote RCM-simulated results.}
    \label{fig5}
\end{figure}

Interpreting CIRBE observations in Fig. \ref{fig1}b and Fig. \ref{fig1}c in relation to specific moments after substorm onset is challenging, making a precise time alignment between the RCM-simulated and CIRBE-observed electron spectra unfeasible. Additionally, discrepancies between the actual and simulated magnetic field configurations prevent an exact point-to-point comparison of magnetic latitude. However, a qualitative comparison between the observed and simulated spectra remains meaningful, as shown in Fig. \ref{fig5}. Since CIRBE observations in Fig. \ref{fig1}b were taken during the early expansion phase (immediately after substorm onset), we selected a corresponding simulation snapshot from the period of sustained electron injection (\(T = 67\) min), specifically Fig. \ref{fig4}g–\ref{fig4}i. To enable a more detailed comparison, we extracted flux-energy profiles along several magnetic latitudes tailward of the POES-determined electron isotropy boundary ($\sim63.27^\circ$), represented by the colored dashed lines in Fig. \ref{fig1}b. A similar approach was applied to the RCM-simulated electron spectrum, where flux-energy profiles were extracted along multiple selected magnetic latitudes, indicated by the colored dashed lines in Fig. \ref{fig4}i.
Due to differences between the simulated and actual magnetic field configurations, which affect the mapping of electron distributions in latitude, it is not possible to extract profiles at exactly the same magnetic latitudes as in the observations. To address this, we carefully selected magnetic latitudes in the simulation that best approximate the observed profiles, ensuring a meaningful comparison despite these inherent discrepancies. The results of this comparison are shown in the four panels of Fig. \ref{fig5}a, where solid lines with circles represent observations, and dashed lines with triangles denote simulation results. For the late expansion phase, corresponding to Fig. \ref{fig1}c, we selected a simulation snapshot a few minutes after the sustained electron injection ended at (\(T = 78\) min), corresponding to Figs. \ref{fig4}j–\ref{fig4}k. The selection of magnetic latitudes follows the same method as before, and the comparison results are presented in Fig. \ref{fig5}b. Across all sub-panels, the observed and simulated spectra exhibit strong consistency, indicating that the RCM simulation effectively reproduces key spectral characteristics.

\begin{figure}[!htbp]
    \centering
    \includegraphics[width=1\linewidth]{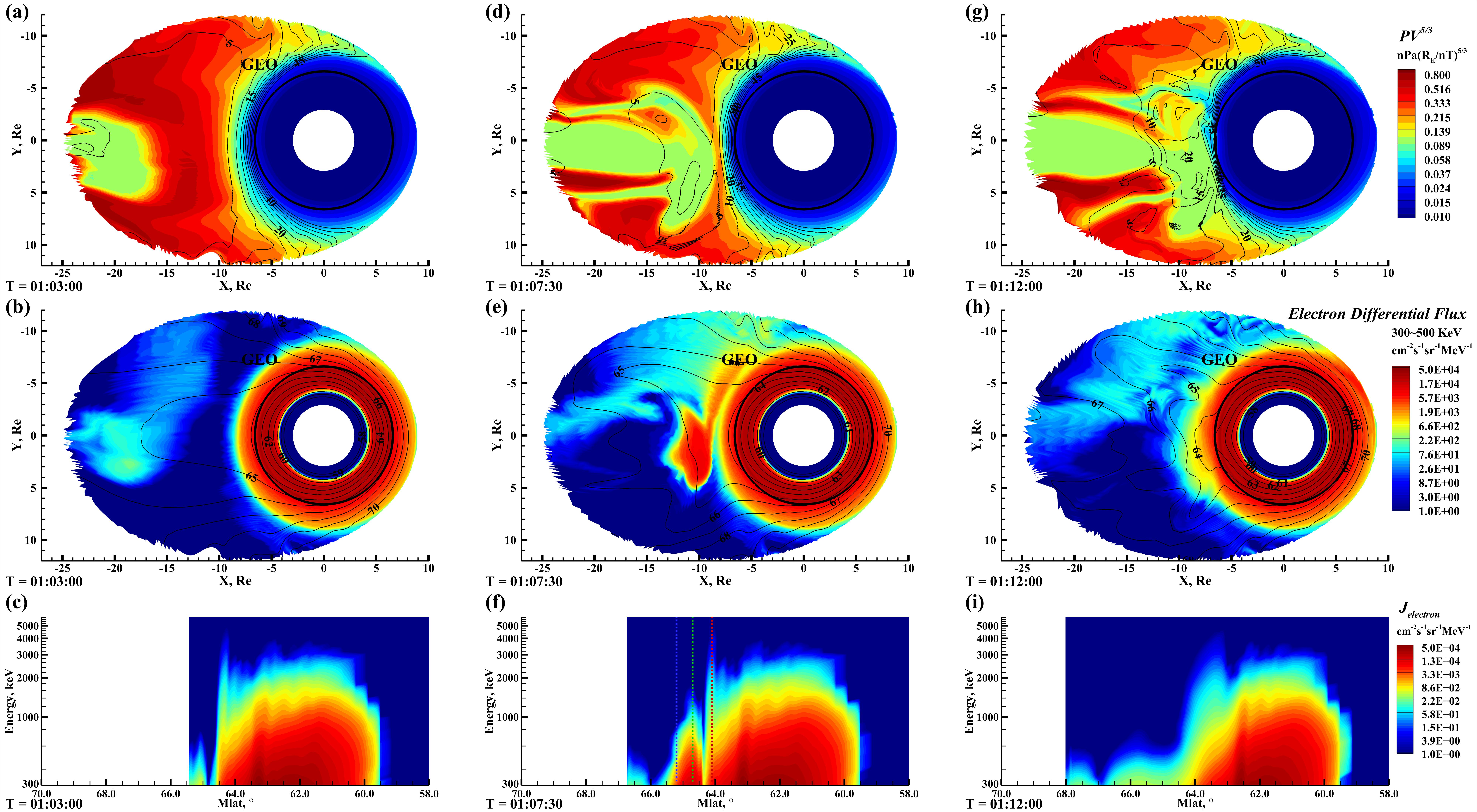}
    \caption{Overview of the RCM expansion phase simulation results for the CIRBE event in Fig. \ref{fig3}, organized in the same format as Figure \ref{fig4}.}
    \label{fig6}
\end{figure}

Figure \ref{fig3} in Section~\ref{sec:data} presents a relativistic electron injection event similar to Figure \ref{fig2}, but with a more distinct spatial separation between the injected population and the outer radiation belt. Given the similarity between these two events, we focus our simulation efforts exclusively on the CIRBE observations in Figure \ref{fig3}. To model this event, we initialized the RCM simulation using the results from the end of the growth phase, as shown in Figs. \ref{fig4}a–\ref{fig4}c. Figures \ref{fig6}a–\ref{fig6}c show the same set of variables as Figs. \ref{fig4}a–\ref{fig4}c, the distribution of flux-tube entropy \(PV^{5/3}\), the 300–500 keV electron differential flux in the RCM equatorial plane, and the energetic electron spectrum along the radial direction at midnight at \(T = 63\) min. To achieve quantitative consistency with the third event in Section~\ref{sec:data}, the flux-tube entropy \(PV^{5/3}\) for this event was only reduced to \( 10\% \) of its pre-onset value at the beginning of the expansion phase. Consequently, the injection channel in Fig. \ref{fig6}a appears noticeably higher in entropy compared to Fig. \ref{fig4}, reflecting the weaker reduction in plasma entropy for this event. Similarly, Figs. \ref{fig6}d–\ref{fig6}f illustrate the propagation of the leading edge of the low-entropy injection channel through the near-Earth magnetotail into the inner magnetosphere, resulting in a significant increase in the 300–500 keV electron differential flux and the formation of a distinct injection-related structure in the electron spectrum of Fig. \ref{fig6}f, which remains clearly separated from the pre-existing high-flux region. Due to the weaker reduction in plasma entropy for this event, the adiabatic heating in this simulation is also less pronounced compared to the first event. As a result, even though Fig. \ref{fig6}f corresponds to a time 30 seconds later than Fig. \ref{fig4}f, the maximum energy of the injected electrons remains lower than that in Fig. \ref{fig4}f. This highlights the dependence of injection-related energy gains on the adiabatic heating. Figures \ref{fig6}g–\ref{fig6}i correspond to the same time record as Figs. \ref{fig4}g–\ref{fig4}i, but with a key difference: the low-entropy bubble in this event only reaches as far as GEO and does not penetrate further inward. This difference is attributed to the extent of \(PV^{5/3}\) reduction, which governs the inward transport of the bubble.  Theoretically, if inertial effects such as overshoot and bouncing are negligible, only bubbles with \(PV^{5/3}\) values within or below the typical GEO value can be injected into the GEO region  \cite{Wolf12}. Once injected, their subsequent motion is expected to be primarily azimuthal, dominated by energy-dependent gradient and curvature drifts, as shown in Fig. \ref{fig4}e and Fig. \ref{fig6}e. Since the \(PV^{5/3}\) reduction in this event is smaller than in the first event, the injection depth of the bubble is correspondingly shallower, as seen in the comparison between Figs. \ref{fig4}g and \ref{fig6}g. Furthermore, because the degree of \(PV^{5/3}\) reduction in the simulation indirectly reflects the extent of magnetic field dipolarization, a smaller reduction indicates a weaker dipolarization. This is evident from the comparison of magnetic field equipotential contours between Figs. \ref{fig4}g and \ref{fig6}g, as well as the expansion of the magnetic latitude range of the high-flux region in Fig. \ref{fig6}h, which is less pronounced than in Fig. \ref{fig4}h, further supporting that this event experienced a weaker dipolarization effect.

\begin{figure}[!htbp]
    \centering
    \includegraphics[width=0.5\linewidth]{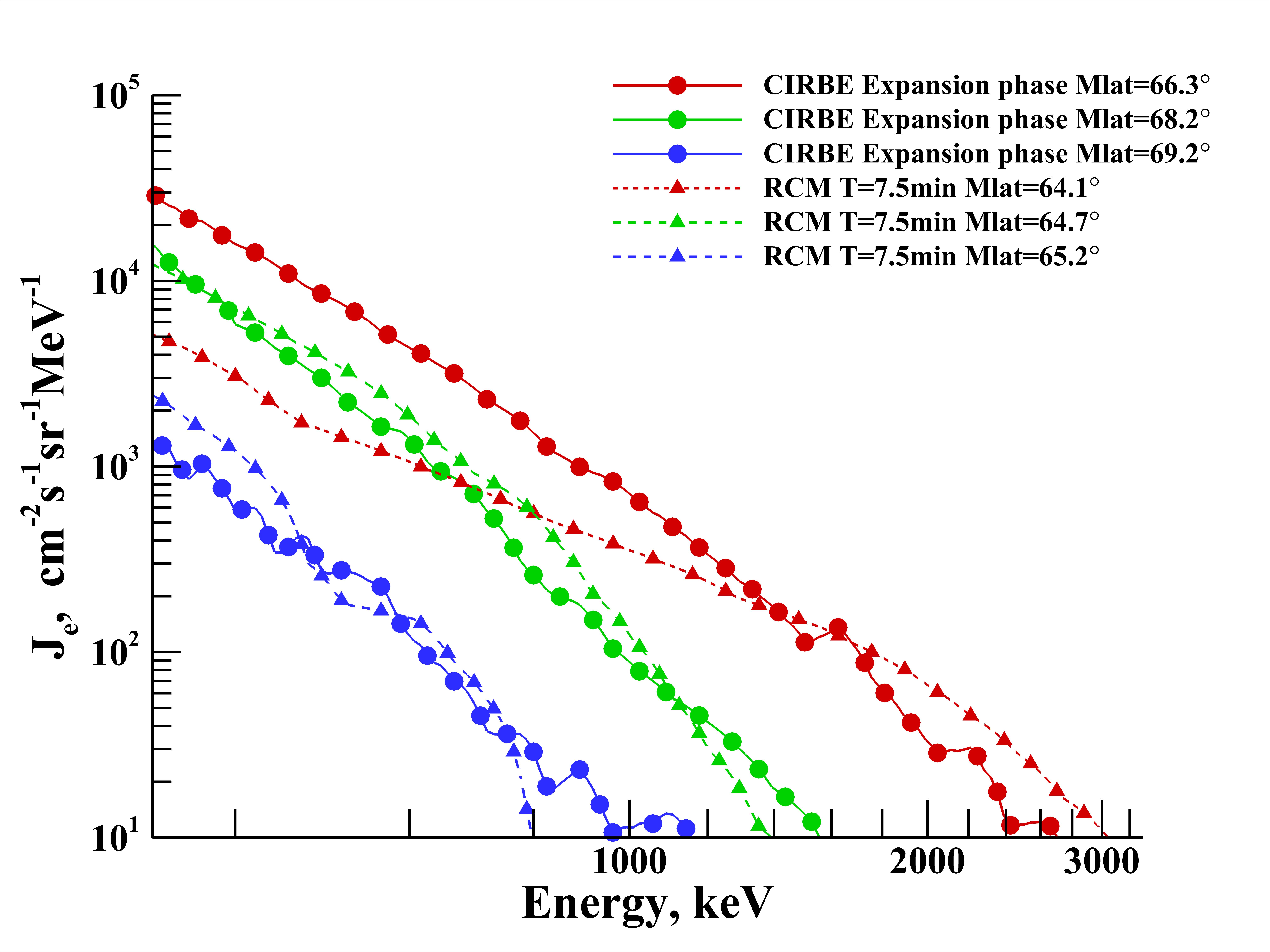}
    \caption{Comparison of the RCM-simulated electron flux with CIRBE observations from Fig. \ref{fig3} during the substorm expansion phase. Solid lines with circles represent CIRBE observations, while dashed lines with triangles denote RCM-simulated results.}
    \label{fig7}
\end{figure}

A qualitative comparison between the CIRBE event in Figure \ref{fig3} and the corresponding simulation results in Figure \ref{fig6} is shown in Figure \ref{fig7}. Since the CIRBE observations in Figure \ref{fig3}c exhibit a distinct, well-separated injection structure, we selected a simulation snapshot where the electron spectrum shows a similarly separated spatial distribution, specifically Figures \ref{fig6}d–\ref{fig6}f.  To enable a more detailed comparison, we extracted flux-energy profiles at three key magnetic latitudes: near the POES-determined electron isotropy boundary ($\sim66.72^\circ$), at the location where the highest-energy injected electrons appear, and at the location closer to the tailward edge of the main injected electron structure. These locations are marked by the red, green, and blue dashed lines in Figure \ref{fig3}c, respectively. A similar method was used to extract flux-energy profiles from the RCM-simulated electron spectrum, with the corresponding colored dashed lines shown in Figure \ref{fig6}f.  The comparison is presented in Figure \ref{fig7}, where solid lines with circles represent CIRBE observations, and dashed lines with triangles indicate the RCM simulation results. In the comparison near the electron isotropy boundary (represented by the red solid and dashed lines), we observe that while the simulated flux curve aligns well with the observed energy range, its flux level at lower energies (a few hundred keV) is notably lower than the observations. This discrepancy primarily arises from the fact that, as indicated by the electron differential flux distribution in Figure \ref{fig6}e and the electron spectrum in Figure \ref{fig6}f, the injected electrons have not yet reached the tailward boundary of the pre-existing high-flux region and thus have not influenced it. This sharp tailward boundary originates from the preceding 1-hour growth phase. Therefore, the difference between the simulation and observations at this location stems from the initial conditions. In other words, if the electron distribution within the corresponding energy and latitude ranges of the RCM had been initialized with a higher flux level at a few hundred keV, where ELFIN observations have currently been used to feed the simulation at \(T = 0\), the simulated result (red dashed line) would be more consistent with the observed result (red solid line). For the other two locations, corresponding to the location where the highest-energy injected electrons appear and the location closer to the tailward edge of the main injected electron structure, the simulated results exhibit good agreements with the observations in both flux level and energy range. This agreement validates the parameter settings used in the simulation. It is worth noting that the magnetic latitude difference between the extracted flux-energy curves from the simulation and the observations is significantly larger in this event compared to Figure \ref{fig5}. This suggests that the actual magnetic field configuration during this event differs substantially from the one used in the simulation. This discrepancy becomes particularly evident when comparing the latitude range of the outer radiation belt during the growth phase of event \#1 in Figure \ref{fig1} and event \#3 in Figure \ref{fig3}, highlighting differences in the magnetic field configuration. As mentioned earlier, our primary focus is on electron energization driven by adiabatic heating, emphasizing how variations in adiabatic heating lead to different levels of electron energization rather than ensuring an exact point-to-point comparison in magnetic latitude. Therefore, the magnetic latitude differences observed in Figure \ref{fig7} are considered acceptable.

\section{Discussion and Conclusions}\label{sec:discussion}
This study leverages coordinated observations from CIRBE and modeling with RCM to explore electron energization mechanisms in the near-Earth magnetotail. The results provide new insights into the origin of relativistic electron populations and demonstrate the effectiveness of combining empirical data and simulation.

First, RCM results validate the approach proposed by \cite{ZhangKun24:ELFIN&CIRBE}, which utilizes POES data to separate plasma sheet and outer radiation belt observations within the CIRBE dataset. This validation is crucial for extending the use of CIRBE measurements to monitor energization processes specifically within the plasma sheet region. By establishing a clear spatial separation between the outer belt and plasma sheet populations, this methodology enables more accurate attribution of energetic electron bursts to local magnetotail dynamics.

Second, CIRBE observations reveal that electron energization within the magnetotail can reach energies of $1\text{--}3$ MeV, significantly exceeding the levels typically observed by near-equatorial missions \cite{Ergun20:observations,Sun22,Sun25,Runov25}. This finding, together with recent reports of $>1$ MeV electron bursts in the magnetotail \cite{Shumko24:arXiv,Zhang25:ELFIN&RX}, highlights the magnetotail as a potential in-situ source of relativistic electron fluxes for the outer radiation belt \cite<see discussion in>{Turner21}. These results challenge the traditional view that such fluxes primarily originate in the outer radiation belt through radial diffusion and local wave-driven acceleration \cite{Green&Kivelson04,Allison&Shprits20,Thorne21:AGU}, suggesting instead that acceleration processes deeper in the tail may play a more significant role than previously appreciated.

Third, the comparison between CIRBE observations and RCM simulations demonstrates that adiabatic heating alone, under the assumption of strong pitch angle scattering and isotropization, can be sufficient to accelerate $<200$ keV electrons to MeV energies within the substorm magnetotail. This supports the idea that large-scale dipolarization and flow braking processes are capable of driving significant energization without necessarily invoking additional non-adiabatic mechanisms. However, the importance of other acceleration processes, particularly those associated with magnetic reconnection, remains an open question. Further investigation is needed to understand the relative contributions of adiabatic and non-adiabatic mechanisms, especially in the context of electron injection, field-aligned acceleration, and resonant wave-particle interactions \cite<see reviews by>{Guo24:ssr,Oka23:ssr,Oka25}.

Despite these promising results of CIRBE/RCM comparison, several limitations should be acknowledged. First, the RCM simulation used here adopts the simplified assumption of isotropization and omits kinetic-scale processes that could alter electron dynamics. Second, the current CIRBE dataset, while providing unique and valuable observations, has limited spatial coverage (in MLT) compared to multisatellite constellations \cite<e.g.,>{Millan&Ukhorskiy24,Ukhorskiy&Millan24}. These factors introduce uncertainties in quantifying the contribution of localized acceleration to global flux enhancements. Finally, although the agreement between simulation and observation supports the plausibility of adiabatic heating, this does not preclude the concurrent role of stochastic or wave-particle processes, which are not explicitly treated in the current modeling framework.

Together, these findings demonstrate the value of CIRBE as a powerful new tool for studying magnetotail electron dynamics, and motivate future studies aimed at resolving the multi-scale and multi-mechanism nature of particle acceleration in Earth's magnetosphere.

\acknowledgments
W.S., X.J.Z., and A.V.A. acknowledge support by NASA contract NAS5-02099 and NSF grant 2400336.

We acknowledge the CIRBE mission team for their efforts in making the dataset accessible to the community, under NASA Heliophysics Division Grants 80NSSC19K0995 and 80NSSC21K0583.

\section*{Open Research}
\noindent CIRBE data are available in CIRBE data archive https://lasp.colorado.edu/cirbe/data-products/ in NetCDF format. THEMIS data are available at http://themis.ssl.berkeley.edu. Data was retrieved and analyzed using PySPEDAS and SPEDAS, see \citeA{Angelopoulos19}.



\end{document}